# Carbon Nanofiber (from fossil fuel) Electric Power Plants:
# Transformation of $CO_2$ Exhaust to Stable, Compact, Valued Commodities

Stuart Licht
Department of Chemistry, George Washington University
Washington, DC USA

**Abstract**
Modes of power plant operation are presented which remove the greenhouse carbon dioxide from fossil fuel plant power station exhausts and transform the carbon dioxide into a valuable carbon nanofiber product. The first mode uses the emissions from a natural gas CC power plant to provide hot $CO_2$ to a molten electrolysis chamber which generates both carbon nanofiber and oxygen. The valuable carbon nanofiber product is removed, heat from the carbon nanofiber and oxygen products is transferred into heating steam for the steam turbine, and the pure oxygen is blended into the air inlet to allow the gas turbine to operate at higher temperature and higher efficiencies. A second mode converts a conventional coal power plant to a STEP coal CNF power plant by directing the hot carbon dioxide combustion emission into carbon nanofiber production electrolysis chamber, and transforming the carbon dioxide to carbon nanofibers with the use of renewable or nuclear energy. Other intermediate modes of fossil fuel carbon nanofiber electric power plants with partial solar input are also evident, as well as a simplified, smaller version (for heating/cooking) rather than electrical production.

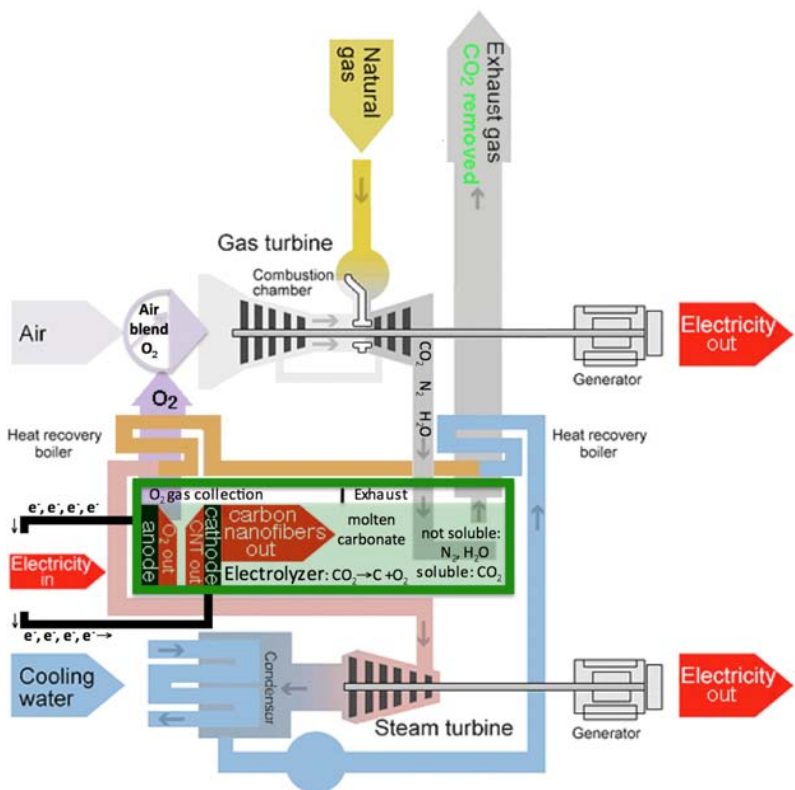

**The Natural Gas Carbon Nanofiber Combined Gas Steam Cycle**

**Carbon Nanofiber (from fossil fuel) Electric Power Plants:**
**Transformation of Carbon Fuel's CO$_2$ Exhaust to Stable, Compact, Valued Commodities**

Stuart Licht
Department of Chemistry, George Washington University
Washington, DC USA

**Introduction**

Fossil fuels comprise approximately 80% of global energy sources and are nonrenewable. When oxidized as fuels, they emit the greenhouse gas CO$_2$, which has risen to an atmospheric concentration of over 400 ppm during the industrial age and contributes to climate change. A recent report on the future of carbon dioxide emission forecasts that from 2010 to 2060 approximately 496 gigatons of CO$_2$ will be generated by fossil fuel combustion.[1] Pathways to avoid this emission are sought (i) to avoid the substantial greenhouse gas climate change consequences and (ii) to preserve this carbon in a compact, energetic form as a resource for the future. We have been developing a high solar energy conversion process, STEP (the solar thermal electrochemical process), which drive the conversion of CO$_2$, and also drives the syntheses of a variety of societal staples without CO$_2$ emissions.[2-9] In this study we probe the cost effective conversion of fossil fuel greenhouse gas emissions to value added commodities.

Recently, we introduced the direct, efficient STEP conversion of atmospheric CO$_2$ to carbon nanofibers. The carbon nanofibers are a valuable commodity due to their unusually high strength, electronic and thermal conductivity and have a wide range of high end applications in construction, transportation, medical, electronics and sports fields. Under the appropriate electrolysis conditions (inexpensive anodes (such as nickel) and cathodes (such as galvanized steel), controlled composition molten carbonate oxide electrolytes, and transition metal growth points, and controlled current densities), carbon dioxide is split into carbon nanofibers and oxygen gas:[10,11]

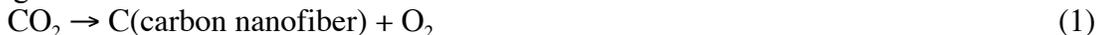

$$CO_2 \rightarrow C(\text{carbon nanofiber}) + O_2 \qquad (1)$$

The chemical and electrochemical conditions of the CO$_2$ transformation are readily controlled, and as shown in Figure 1, under the correct conditions can generate either straight or tangled carbon nanofibers and a carbon nanotubes subset of carbon nanofibers, carbon nanotubes.[10,11] The electrochemical energy required to form the carbon tubes is as low as 0.8V and carbon nanofibers can be formed at a high rate (rates of 0.1 to 1 amps per cm$^{-2}$ of electrode) at electrochemical potentials over 1V.[12]

In this study we develop processes for the transformation of carbon dioxide to carbon nanofibers from more concentrated CO$_2$ sources, industrial flue gas, and in particular to remove the carbon dioxide from fossil fuel electric plants while producing the valuable carbon nanofiber product. This provides less of a challenge than the previously demonstrated direct removal of atmospheric carbon dioxide, which is dilute (0.04%), as the flue gas is already hot, and several hundred fold more concentrated in CO$_2$. CO$_2$ is readily soluble in the molten carbonate oxide electrolytes, while nitrogen and water vapor are not. The undissolved gases (N$_2$ and H$_2$) exhaust from the electrolyzer with the CO$_2$ removed (and transformed to carbon nanofibers). If not removed in pre-exhuast states, lower levels of impurities, such as sulfur gases are reduced to sulfur in the electrolysis chamber[13] and do not inhibit or poison the electrolysis process. The dissolution of carbon dioxide in molten carbonates occurs via reaction with the dissolved oxide, for example in the case of lithium carbonate:

$$CO_2 (gas) + Li_2O(dissolved) \rightarrow Li_2CO_3(molten) \quad (3)$$

Electrolysis with 4 electrons per molecule yields:

$$Li_2CO_3 \rightarrow C(carbon\ nanofiber) + Li_2O + O_2 \quad (4)$$

Note, that the net reaction of equations 3 and 4 equals equation 1.

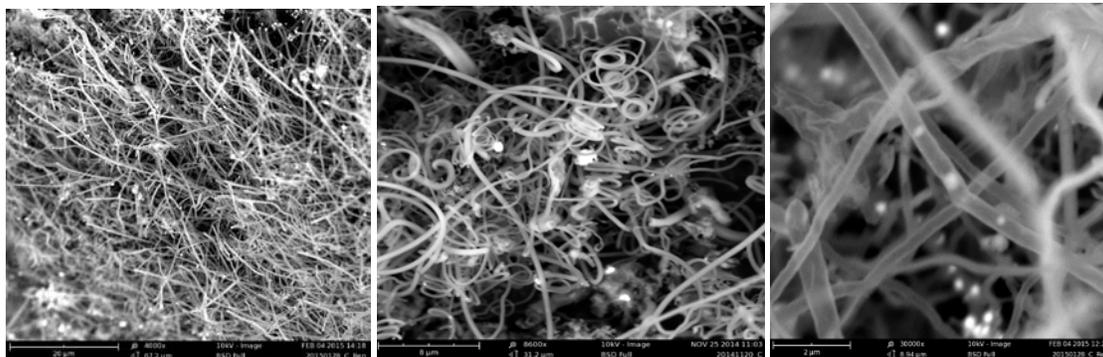

**Figure 1**. Carbon dioxide transformed into uniform straight (left) and tangled (middle) carbon nanofibers[10] and carbon nanotubes[11] (right).

In this study we probe efficient designs for the conversion of (i) CC (combined cycle) natural gas electric power plants and (2) coal plants to (1) CC CNF (combined cycle carbon nanofiber) and (2) STEP coal CNF (STEP Coal carbon nanofiber) electric power plants.

**1. Natural Gas Combined Cycle Carbon Nanofiber Power Plant.**
**1a. Natural gas fuels.**
The composition of natural gas used to drive combined cycle (CC) power plants varies widely, but generally contains approximately 95% methane ($CH_4$), 3% ethane ($C_2H_6$), 1% nitrogen, 0.5% carbon dioxide ($CO_2$), 0.2% propane ($C_3H_8$), and less than 0.1% of other organics (primarily alkanes), oxygen ($O_2$) and sulfur compounds, although the ratio of methane to other organics can vary widely.[14]

**1b. Exhaust from conventional CC gas electric power stations.**
Conventional coal CC (gas/steam turbine combined cycle) electric power stations emit massive amounts of carbon dioxide to the atmosphere. Combined cycle gas power plants energetically operate more efficiently (~55%) than coal power plants (~35%) and exhaust less carbon dioxide per watt of power. Exhaust flue gas volume composition varies with plant construction. The flue gas volume is ~295 $m^3$/GJ respectively from gas CC power plants. The gas contains a majority of nitrogen, water vapor, and ~9% or ~14% $CO_2$ in the flue gas respectively from gas or coal power plants.[15] Depending on the source and purity of the natural gas, additional infrastructure can be included to scrub the flue gases of sulfur, nitrous oxides and heavy metals.

## 1c. Conventional Combined cycle CC Gas/Steam Power plant

The left side of Figure 1 illustrates the action of a CC electrical power plant which emits a flue gas that contains ~9% carbon dioxide (compare to 0.04% $CO_2$ in ambient air). As illustrated, the conventional CC plant increases the fuel to electricity efficiency compared to single cycle electrical power plants by directing the exhaust heat emissions from a gas turbine (the Brayton thermodynamic cycle to generate electricity) to boil water to power a steam turbine (the Rankine thermodynamic cycle to generate electricity).

In this CC power plant, the heat available from regular (25°C) combustion is:
$$CH_4 + 2O_2 \rightarrow CO_2 + 2H_2O \qquad\qquad 889 \text{ kJ/mole} \qquad (5)$$

and approximately 55% of this heat is converted to combined electric power by the gas (top) and steam (bottom) generators of Figure 2 left.

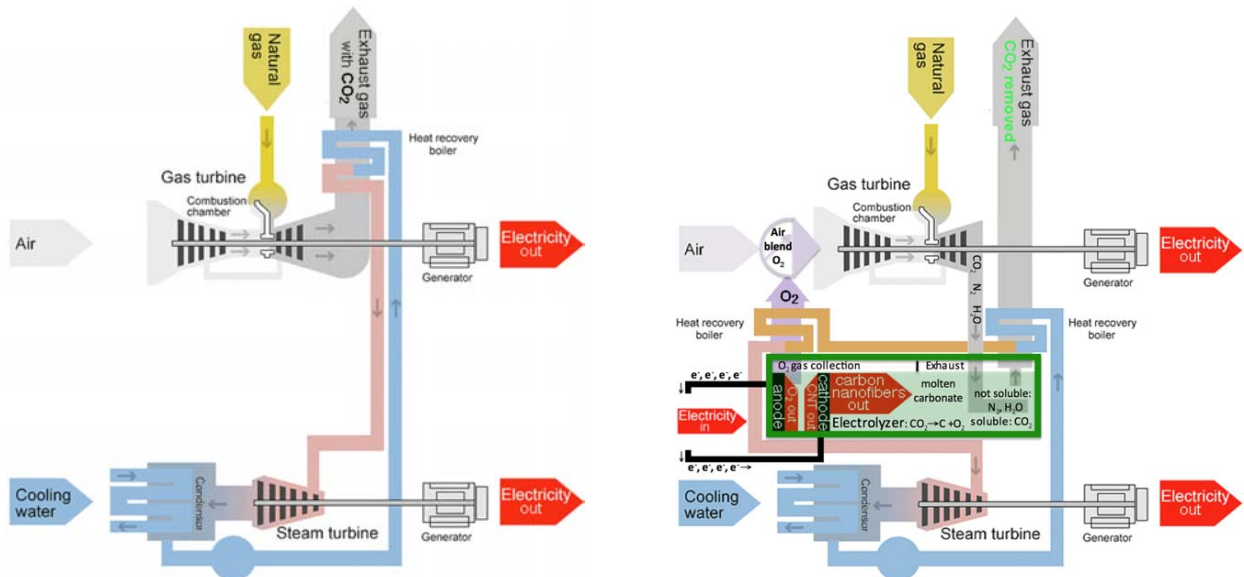

**Figure 2**. The action of a conventional CC power plant which has an exhaust with $CO_2$ (left) compared to the introduced CNF CC power plant (left) which has carbon dioxide removed from the exhaust gas (right) and instead produces a valuable carbon product. Left: The conventional gas/steam combined cycle CC power plant is illustrated as modified from reference 16. Right: The CC CNF power plant in this study, illustrating the CNF (including carbon nanofibers or carbon nanotube) product, the electrolysis pure oxygen cycled back into the gas combustion increasing efficiency, the recovered heat returned to the steam turbine, and the carbon dioxide removed from the exhaust gas.

## 1d. CC CNF Power Plant: The Natural Gas Carbon Nanofiber Combined Gas Steam Cycle

As shown on the right side of Figure 2, the **Gas** Combined Cycle Carbon Nanofiber **or** CNF CC plant product, in addition to electricity, simultaneously produces a valuable carbon nanofiber product. Unlike the conventional CC on the left side of the figure, $CO_2$ is removed from the exhaust of the CNF CC power plant on the right. As shown, the hot $CO_2$, $N_2$ and $H_2O$ CC emission is instead bubbled into molten carbonate where only the $CO_2$ dissolves. The remaining gases (with the carbon dioxide removed) become the exhaust gas (after heat recovery). The dissolved carbon dioxide is split by electrolysis into oxygen gas at the anode, which is fed (after heat recovery) back into the gas turbine and carbon (CNF) at the cathode. The CNF product is tuned for strength, diameter, length, vacancy, geometry, and electrical or thermal conductivity by the specific molten salt electrochemistry employed. The CNF product may be removed periodically or as a constant throughput). The recovered heat boils water to power a steam turbine to also generate electricity. Heat is returned to the steam generator chamber using (i) heat recovered from the electrolysis (pure oxygen and carbon nanofiber) products, and (ii) from the carbon dioxide removed electrolysis exhaust.

The CNF CC natural gas to electricity plant efficiency is increased by the "free" pure oxygen generated during the carbon nanofiber electrolysis and is offset by the energy required to drive the electrolysis. The minimum energy required to drive the electrolysis is very low (<< 1 V[12]) and increases with production rate. This is recovered many fold in value by the increased value of the CNF product compared to the cost of the natural gas consumed. In 1 atm 20°C air, a methane flame temperature varies from 900 to 1,500 °C, and the temperature increases with combustion in pure $O_2$, rather than air.

Oxy-fuel combustion has been used to increase fuel efficiency (and increase $CO_2$ concentration for conventional sequestration) in IGCC electric power plants (section 2). Here this pure $O_2$ is utilized without the energy loss of the cryogenic "cold box" of the bottom panel of Figure 3, or other forms of oxygen separation from air. Increased efficiency is balanced with materials constraints of too high a temperature by blending $O_2$ concentrations higher than that in air, but lower than pure $O_2$. In the new CNF CC design presented here, to a first order approximation the increased efficiency from the blended pure oxygen offsets the consumed electrolysis, such that the CNF CC power plant retains at a minimum the approximate electrical energy efficiency as a conventional coal power plant (> 35%) with further increases in efficiency as the CNF CC system advances.

**2. Coal Nanofiber Power Plant.**
**2a. Coal fuel.**
Coal is principally carbon and moisture, and natural gas is principally composed of methane. More specifically for the coals lignite contains 24-35% carbon and up to 66% moisture, bituminous coal contains 60-80% carbon, while anthracite is 92 to 98% carbon. The three respectively have heat contents of ~15, 24 to 35, and 36 kJ/g.[17]

**2b. Exhaust from conventional coal power stations.**
Conventional coal electric power stations emit massive amounts of carbon dioxide to the atmosphere. Exhaust flue gas volume composition varies with plant construction. The flue gas volume is ~ ~323 $m^3$/GJ from coal power plants. The flue gas contains a majority of nitrogen, water vapor, and ~14% $CO_2$.[15] Additional infrastructure is included to scrub the flue gases of sulfur, nitrous oxides and heavy metals.

**2c. Coal Power plants.**
The vast majority of coal power plants simply combust pulverized coal to produce steam, which drives turbines to generate electricity. The exhaust of the combustion is generally cleansed of the majority of sulfur, nitrogen, heavy metal and particulates and the remaining flue gas exhaust, which contains a high carbon dioxide content (along with nitrogen and water vapor) is emitted directly to the atmosphere.[6]

There are a relatively few number of integrated (coal) gasification combined cycle (IGCC) power plants, which gasify the coal to hydrogen. An IGCC is illustrated on the bottom panel of Figure 2.[7] The heat from combustion of the gas produced from the coal drives a turbine(s), while heat recovered from this combustion, as well heat as from the coal gasification process, boils water to drive a second set of turbines. These IGCC plants higher energy conversion efficiency (~50% compared to ~35% for traditional), but have higher capital costs related to the increased complexity and higher grade materials need of the increased temperature combustion. The coal gasification generates a more concentrated carbon dioxide emission than simple coal combustion for heat. Coal gasification proceeds from coal combustion with a lean oxygen mix, to carbon monoxide (CO; with $CO_2$), to syngas (CO + $H_2$; with $CO_2$), and/or to $H_2$ (via the water shift reaction, $CO + H_2O \rightarrow CO_2 + H_2$).[8]

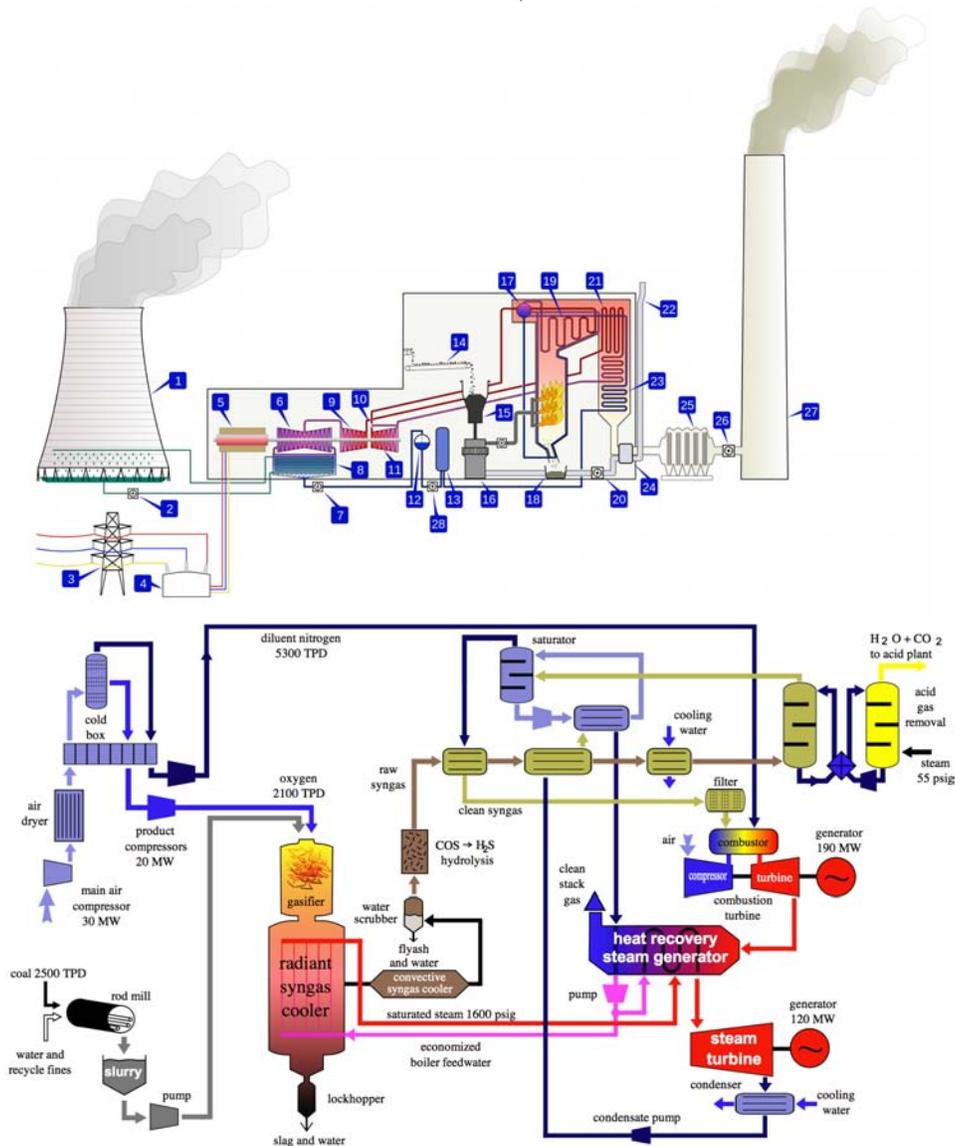

**Figure 3.** Illustrative examples of coal power plants. Conventional (top)[6] and integrated gasification combined cycle (IGCC, bottom)[8] coal power plants. Key for conventional power plant: 1. Cooling tower, 2. Cooling water pump, 3. Transmission line (3-phase), 4. Step-up transformer (3-phase), 5. Electrical generator (3-phase), 6. Low pressure steam turbine, 7. Condensate pump, 8. Surface condenser, 9. Intermediate pressure steam turbine, 10. Steam Control valve, 11. High pressure steam turbine, 12. Deaerator, 13. Feedwater heater, 14. Coal conveyor, 15. Coal hopper, 16. Coal pulverizer, 17. Boiler steam drum, 18. Bottom ash hopper, 19. Superheater, 20. Forced draught (draft) fan, 21. Reheater 22. Combustion air intake, 23. Economiser, 24. Air preheater, 25. Precipitator, 26. Induced draught (draft) fan, 27. Flue-gas stack.

**2d. Alternate modes of introducing added heat in thermal and solar heat to coal plants.**
In a manner similar to that described here for the conversion of natural gas CC power plants to CC CNF power plants, the IGCC can be converted to IGCC CNF power plants, with the additional opportunity that the initial lean oxygen coal gasification step prefers an enriched oxygen (nitrogen deplete) source such as is available from the CNF electrolysis anode product output. Also, in a manner similar to that described for the conversion of CC power plants to CC CNFplants, there are several points in a conventional, pulverized coal electric power plant (as illustrated on the top panel of Figure 3), in which heat may be extracted for addition of an $CO_2$ to CNF electrolysis chamber. These permutations to both the conventional coal and IGCC power plants are evident from Figure 1, and will not be detailed in this study. Instead the design of the conventional coal power plants will be extended, rather than substantially modified, to produce carbon nanofibers using solar energy.

Solar thermal energy can be incorporated at several points in the heat exchange processes in the top of Figure 2. However, rather than this, an expeditious path to incorporate solar energy and remove the $CO_2$ from the coal plant exhaust is accomplished without substantially changing the existing coal plant infrastructure by direct coupling of the solar process to the existing coal plant exhaust. Figure 4 summarizes key components of a coal power plant from the top of Figure 3, including the combustion of the pulverized coal fuel with air, the extraction of combustion heat to drive a generator and the exhaust emitted to the atmosphere containing nitrogen, carbon dioxide and water vapor.

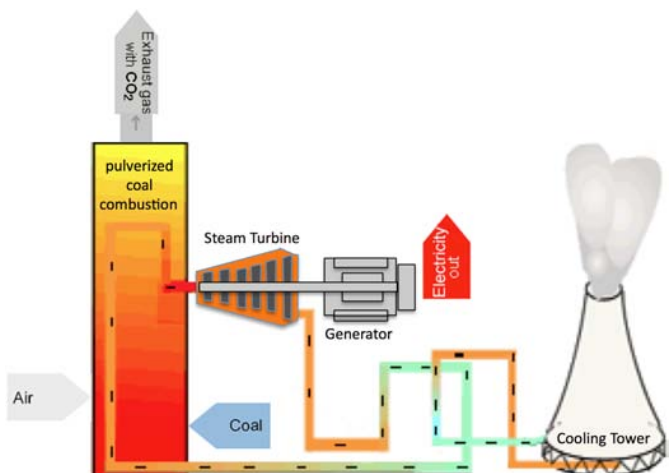

**Figure 4**. Simplified illustration of a coal power plant.

Figure 5 illustrates various paths by which STEP (the solar thermal electrochemical process) can utilize solar thermal energy, including on the left side, by exchange of excess heat from a photovoltaic to the STEP reactants. A second optic mode is shown in the middle panel via a solar tower in which mirrors, mounted on heliostats to track the sun, reflect and concentrate the sunlight at a central point, where upon it is split into the visible concentrated sunlight (to drive concentrator photovoltaics providing electronic charge to the electrolyzer) and thermal sunlight (providing heat to the heat the electrolyzer). Finally a third optic mode is accomplished via short path length solar concentrators such as parabolic mirrors, or convex or Fresnel lenses with the light split into infrared and visible components via a dielectric filter (also referred to as a hot/cold mirror). The wavelength at which light is reflected or transmitted through the filter is tuned by the dielectric.[9] The right side design will be used to symbolize the various alternative STEP optic modes in Figure 6.

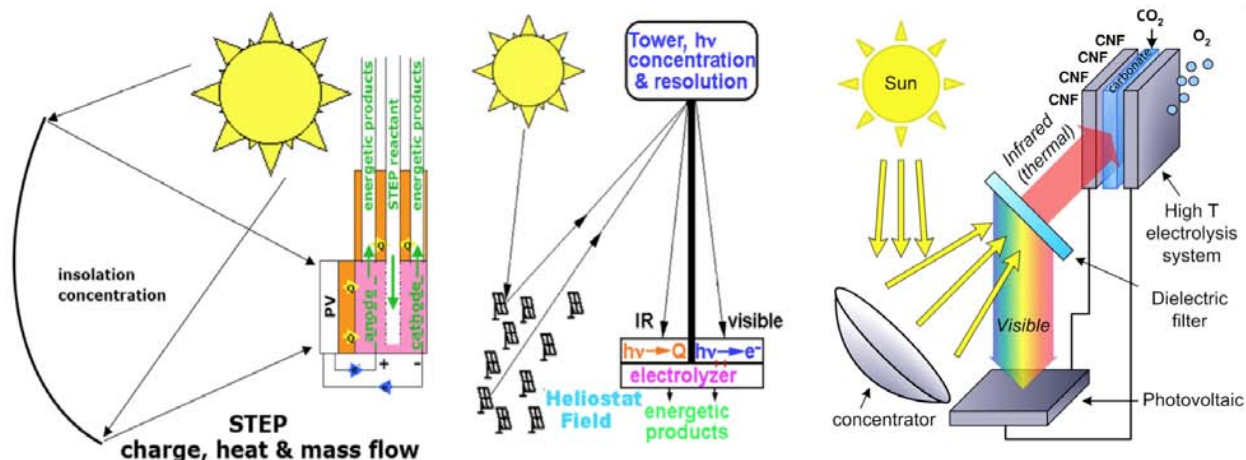

**Figure 5**. There are various optical modes of solar concentration available to heat the electrolyte in STEP including via excess heat transferred from the photovoltaic (left) or via splitting concentrated sunlight into separate solar visible and solar IR (thermal) energy via long (middle) or short (right) focal length optics.

## 2e. The STEP Coal/CNF Power Plant.

Figure 6 illustrates a coal electric plant in which carbon dioxide has been removed from the exhaust, and provides a carbon source to generate a value-added carbon nanofiber product using the solar thermal electrochemical process. In this STEP coal/CNF the heat is extracted from the oxygen product and CNF products to decrease the heat needed in the electrolysis chamber, and pure oxygen product is mixed with the air inlet to increase efficiency of the coal combustion and electric power generation. Specifically, two modes of STEP are illustrated in the figure and both use concentrated sunlight. In STEP, sunlight is split into its visible and infrared spectra. The visible sunlight drives an efficient concentrator photovoltaic (or solar cell), while the thermal (infrared) sunlight heats the electrolysis chambers.[9] directed to heat the electrolysis chamber, and any form of renewable or nuclear electric energy is drive the electrolysis transforming $CO_2$ to CNF (and $O_2$). In Hy-STEP (hybrid-STEP), all sunlight is directed to heat the electrolysis chamber, and any form of renewable or nuclear electric energy is drive the electrolysis transforming $CO_2$ to CNF (and $O_2$).[10,11,12]

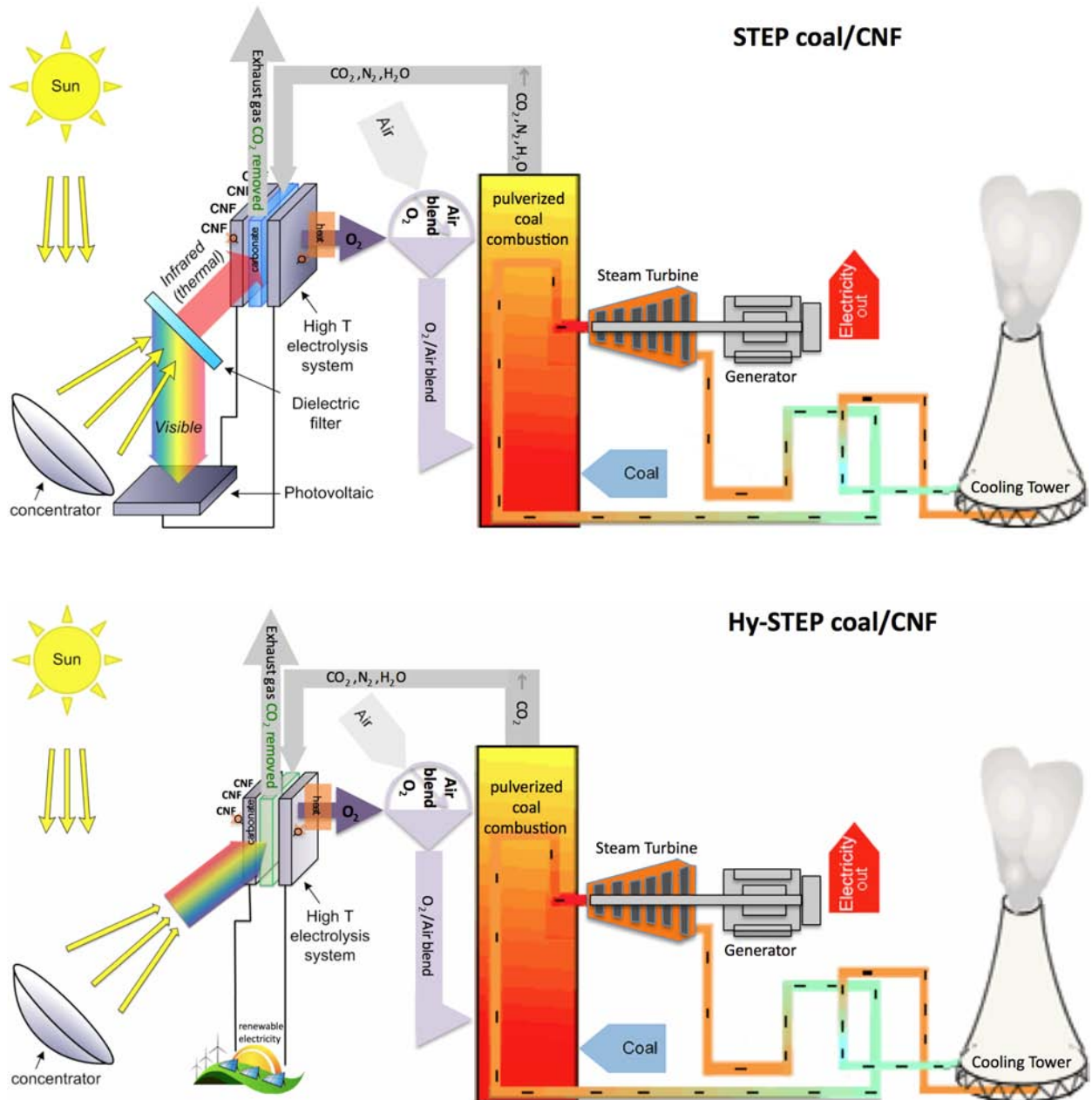

**Figure 6**. Two modes of STEP CNF in which the coal power plant $CO_2$ emission is transformed into a valued added carbon nanofiber commodity. In STEP coal/CNF (top panel), concentrated sunlight is split into two band, the infrared band heats the electrolysis chamber, and the visible all the concentrated sunlight is directed to heat the electrolysis chamber, and driven by any form of renewable or nuclear generated electricity.

## 3. Conclusions.

Valuable carbon nanofiber, CNF, products replace the $CO_2$ emission of power plants. The synthesis of CNFs has been of increasing interest, with applications ranging from capacitors and nanoelectronics, Li-ion batteries and electrocatalysts to the principal component of lightweight, high strength building materials.[18,19] CNFs formed from $CO_2$, can contribute to lower greenhouse gases for example by consuming, rather than emitting $CO_2$, and by providing a carbon composite material that can be used as an alternative to steel, aluminum, and cement whose productions are associated with massive $CO_2$ emissions.[4-8] Carbon composites will further decrease emissions by facilitating both wind turbines and lightweight, low-carbon-footprint transportation.[20]

Modes of power plant operation are presented which remove the greenhouse carbon dioxide from fossil fuel plant power station exhausts and transform the carbon dioxide into a valuable carbon nanofiber product. The first mode uses the emissions from a natural gas CC power plant to provide hot $CO_2$ to a molten electrolysis chamber which generates both carbon nanofiber and oxygen. The valuable carbon nanofiber product is removed, heat from the carbon nanofiber and oxygen products is transferred into heating steam for the steam turbine, and the pure oxygen is blended into the air inlet to allow the gas turbine to operate at higher temperature and higher efficiencies. A second mode converts a conventional coal power plant to a STEP coal CNF power plant by directing the hot carbon dioxide combustion emission into carbon nanofiber production electrolysis chamber, and transforming the carbon dioxide to carbon nanofibers with the use of renewable or nuclear energy. Intermediate modes in which the sunlight is split to generated both electrical (from visible) and thermal (from infrared) sunlight is shown as well as a mode in (Hy-STEP) in which all sunlight is directed to heat the electrolysis chamber. Other intermediate modes of fossil fuel carbon nanofiber electric power plants with partial solar input are also evident. A simplified, smaller version (not illustrated, for heating/cooking) rather than electrical production, would place the carbon nanofiber production electrolysis chamber in the exhaust path of a fossil or biofuel stove, with renewable energy to generate electricity for the electrolysis and returning the hot oxygen to the fuel chamber to facilitate higher efficiencies and more complete combustions.

# Title:
# Carbon nanofibers, precious commodities from sunlight & $CO_2$ to ameliorate global warming

Stuart Licht, Jiawen Ren
Department of Chemistry, George Washington University
Washington, DC, 20052, USA.

This study introduces the high yield, electrolytic synthesis of carbon nanofibers, CNFs, directly from carbon dioxide. Production of a precious commodity such as CNFs from atmospheric carbon dioxide provides impetus to limit this greenhouse gas and mitigate the rate of climate change. CNFs are formed at high rate using inexpensive nickel and steel electrodes in molten electrolytes. The process is demonstrated as a scaled-up stand-alone electrolytic cell, and is also shown compatible with the STEP, solar thermal electrochemical process, using concentrated sunlight at high solar to electric efficiency to provide the heat and electrical energy to drive the CNF production.

Anthropogenic climate change consequences will be resolved when we inexpensively convert atmospheric carbon dioxide to diamonds yielding a compact, stable repository of $CO_2$ as valuable carbon. The global warming consequences of increasing atmospheric carbon dioxide concentrations are well established[1,2] as glacier and ice cap loss, sea level rise, droughts, hurricanes, species extinction, and economic loss. Here, we show another highly valued, stable, compact carbon, carbon nanofiber, CNF (not diamonds), that can be directly formed from atmospheric or exhaust $CO_2$ in an inexpensive process. Whereas the value per tonne of coal is ~$60 and graphite is ~$1,000, carbon nanofiber ranges from $20,000 to $100,000[3] per tonne. Today, CNFs require 30 to 100 fold higher production energy compared to aluminum[4]. The "production of CNTs and nano-fibers by electrolysis in molten lithium carbonate is impossible" according to the prior literature[5]. We present STEP (solar thermal electrochemical process) carbon nanofibers, in which solar energy efficiently drives the direct electrolytic conversion of $CO_2$, dissolved in molten carbonates to CNFs at high rates using scalable, inexpensive nickel and steel electrodes. The structure is tuned by controlling the electrolysis conditions, such as the addition of trace nickel to act as CNF nucleation sites. We calculate that with an area <10% the size of the Sahara, the new STEP CNF can remove and decrease atmospheric $CO_2$ to pre-industrial revolution levels in ten years. New infrastructure and merchandise built from CNFs would provide a massive repository to store atmospheric $CO_2$. CNFs are increasingly used in high strength, composite building materials ranging from top-end sports equipment to lightweight car and airplane bodies.[3] CNFs not only bind $CO_2$, but will eliminate the massive $CO_2$ emissions associated with the production of steel, aluminum, and cement[6-9], and will further decrease emissions by facilitating both wind turbines and lightweight, low carbon footprint transportation[10].

Prior to the recognition of a variety of unique carbon nanoscopic structures such as fullerenes, nanotubes, and nanofibers starting in 1985[11], the reduction of carbonates to (macroscopic) carbons in inorganic molten electrolytes from hydroxides and a barium chloride/ barium carbonate melt was recognized as early as the late 1800's[12]. Today, the principal methods of CNF preparation are (i) spinning of polymer nanofiber precursors followed by carbonization heat treatment[13-15] and (ii) catalytic thermal chemical vapor deposition (CVD) growth[13,14]. CVD growth of CNFs is catalyzed by metals or alloys, which are able to dissolve carbon to form metal carbides including nickel, iron, cobalt, chromium, and vanadium[13,14,16]. Low levels of the catalyst relative to CNF are required, and prominent catalyst nanoparticles are observed at the fiber tip, as well as catalyst clusters along the fiber that has migrated during CNF growth[16]. A variety of CNF morphologies have been observed including linear, coil, and sphere clustered CNFs[13-18].

The electrochemical synthesis of CNFs has not been widely explored. Solid carbon electrodes have been electrolytically converted to nanostructures such as nanotubes in molten halide solutions via alkali metal formation, intercalation into, and exfoliation of the carbon[19,20]. Instead of the conversion of solid carbon, the rate of the direct reduction of $CO_2$ studied with carbon and platinum electrodes is limited by the low solubility of $CO_2$ in molten halides requiring high (15 atm) $CO_2$ pressure, and is accompanied by corrosion of the electrodes[20,21]. A study of 5-10% $Li_2CO_3$ in molten chloride concluded that "production of CNTs and nano-fibers by electrolysis in molten lithium carbonate is impossible" because "Reduction and deposition of carbon occur instead of lithium discharge and intercalation into the cathode.[5]" That is correct, but did not anticipate alternative CNF Ni nuclei growth paths from molten $Li_2CO_3$ shown here.

We have introduced STEP (the <u>s</u>olar <u>t</u>hermal <u>e</u>lectrochemical <u>p</u>rocess), an alternative solar energy conversion process that uses the full spectrum of sunlight to drive solar syntheses of a range of energetic molecules at high solar efficiency. Solar thermal, high reactant concentrations, and unusual chemistries are used to decrease the energy required to drive endothermic electrolyses, and renewable electrical current is used to drive the low energy electrolysis[22]. STEP products are energetic molecules rather than electricity, and solar conversion efficiencies of 34 to 54% have been achieved[6,23]. STEP carbon[6,23,24], STEP iron[6-8,25], STEP organic[26], STEP cement[9], STEP Fuels[27], and STEP ammonia[28,29] are among the processes which have been demonstrated. Here, we show the first demonstration in which STEP carbon is capable of generating pure carbon nanofibers from atmospheric carbon dioxide using molten carbonate electrolytes. Molten carbonates, such as pure $Li_2CO_3$ (mp 723°C) or lower melting point carbonate eutectics such as $LiNaKCO_3$ (mp 399°C) or $LiBaCaCO_3$ (mp 620°C), mixed with highly soluble oxides, such $Li_2O$ and $BaO$, sustain rapid absorption of $CO_2$ from the atmospheric exhaust $CO_2$. Equilibrium constraining lithium or lithium/barium oxide absorption has been presented, and the lithium case is described as[6,8,24]:

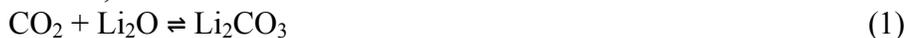
$$CO_2 + Li_2O \rightleftharpoons Li_2CO_3 \tag{1}$$

Air contains 0.04% $CO_2$; this is only $1.7 \times 10^{-5}$ mol of tetravalent carbon per liter, whereas molten carbonate contains ~20 mol of reducible tetravalent carbon per liter. A separate process to concentrate atmospheric $CO_2$ is not needed in STEP carbon. Hence, by absorbing $CO_2$ from the air, molten carbonates provide a million-fold concentration increase of reducible tetravalent carbon available for splitting (to carbon) in the electrolysis chamber. Carbonate's higher concentration of active, reducible tetravalent carbon sites logarithmically decreases the electrolysis potential and can facilitate charge transfer at low electrolysis potentials. $CO_2$ is bubbled into the molten carbonate, and during electrolysis, oxygen is evolved at the anode while a thick solid carbon builds at the cathode (Figure 1). We observe that carbonate is readily split to carbon approaching 100% coulombic efficiency (coulombic efficiency is determined by comparing the moles of applied charge to the moles of product formed, where each mole of product formed depends on four moles of electrons) at molten carbonate temperatures of 750°C and below[6,9,23,24]; this is the case unless hydroxides are mixed with the carbonate, whereupon $H_2$ and carbon products are co-generated[27]. High current densities (>1 amp $cm^{-2}$) of carbon formation[5] are sustained, and we observe similar sustained currents at carbon, platinum, or steel cathodes (the latter effectively become carbon electrodes during the deposition). Note via the Faraday equivalence that 1 A $cm^{-2}$ will remove 36 tonnes of carbon dioxide per $m^2$ cathode per year. Full cell electrolysis potentials range from ~1V under conditions of high temperature (*e.g.* 800°C), low current density (*e.g.* 10 mA $cm^{-2}$), and high oxide concentration (*e.g.* 6 molal $Li_2O$), to several volts. Conditions that increase carbonate electrolysis voltage are high current density, lower temperature, or lower oxide concentration.

At higher temperatures, the product gradually shifts to a mix of carbon and carbon monoxide, and it becomes pure CO by 950°C[23]. The respective 4- or 2-electron processes are given by:

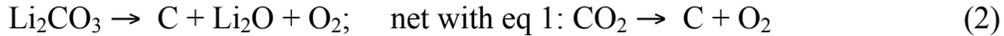

$Li_2CO_3 \rightarrow C + Li_2O + O_2$; net with eq 1: $CO_2 \rightarrow C + O_2$ (2)

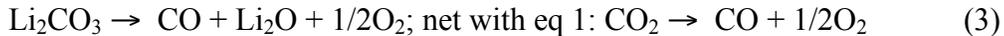

$Li_2CO_3 \rightarrow CO + Li_2O + 1/2O_2$; net with eq 1: $CO_2 \rightarrow CO + 1/2O_2$ (3)

Electrolysis, either via equation 2 or equation 3, releases $Li_2O$ to permit continued absorption of carbon dioxide (net: $CO_2$ is split and oxygen is released).

The SEM in Fig. 1 evidences no CNF in the carbon product deposited at the cathode subsequent to carbonate electrolysis in a nickel-free environment ($Li_2CO_3$ at 730°C with 6 molal $Li_2O$ in the absence of nickel, utilizing a Pt rather than a Ni anode). Amorphous and platelet structures are seen, with the platelets indicative of partially formed multi-layered graphene/graphite. Electron dispersive spectroscopy elemental analysis indicates that the amorphous and platelet structures are composed of > 99% carbon. Similarly, CNF formation was not observed in the cathode product when the electrolysis was instead conducted with a Ni anode in a corrosion-free lower temperature (630°C) $Li_{1.6} Ba_{0.3}Ca_{0.1}CO_3$ electrolyte.

We had not previously anticipated the oxygen generating anode effects on the structure of the carbon formed at the cathode during carbonate electrolysis. As demonstrated here, these anode effects are highly specific and can promote significant carbon nanofiber formation. We have investigated Pt, Ir, and Ni, and each can be effective as oxygen generating anodes[6,8,23]. Whereas Ir exhibits no corrosion following hundreds of hours use in molten carbonates, the extent of Ni corrosion is determined by the cation composition of the carbonate electrolyte. A nickel anode undergoes continuous corrosion in a sodium and potassium carbonate electrolyte[27], it is stable after initial minor corrosion in lithium carbonate electrolytes[8], and no corrosion of the nickel anode is evident in barium/lithium carbonate electrolytes[24,27]. In lithium carbonate electrolytes, we have quantified the low rate of nickel corrosion at the anode as a function of anode current density, electrolysis time, temperature, and lithium oxide concentration[9]. The Ni loss at a 100 mA cm$^{-2}$ Ni anode in Li$_2$CO$_3$ at 750°C with 0 or 5 molal added Li$_2$O is respectively 0.5 or 4.1 mg cm$^{-2}$ of anode subsequent to 600 seconds of electrolysis, and increases to 4.6 or 5.0 mg cm$^{-2}$ subsequent to 1200 or 5400 seconds of electrolysis. The Ni loss increases to 7.0 mg or 13.8 mg cm$^{-2}$ respectively subject to higher current (1000 mA cm$^{-2}$) or higher temperature (950°C). Each of these nickel losses tends to be negligible compared to the mass of Ni used in the various Ni wire or Ni shim configured anodes. Nickel oxide has a low solubility of 10$^{-5}$ moles NiO per mole of molten Li$_2$CO$_3$[30], equivalent to 10 mg Ni per kg Li$_2$CO$_3$. This low, limiting solubility constrains some of the corroded nickel to the anode surface as a thin oxide overlayer, with the remainder as soluble oxidized nickel available for reduction and redeposition at the cathode.

Electrolyte composition and current density substantially alter the nature of metal electrodeposition from molten carbonates. This is easy to observe with a more soluble metal salt, as opposed to one that is less soluble. We have recently demonstrated that iron (oxides), while insoluble in sodium or potassium carbonates, can become extremely soluble in lithiated molten carbonates, with solubility increasing with temperature to over 20% by mass Fe(III) or Fe(II) oxide[6-8,25]. The Extended Data Fig. 1 demonstrates that the size of deposited Fe varies inversely with current density from molten carbonates at a fixed charge (1Ah over a 10 cm$^2$ cathode), and that the iron shape is strongly affected by the electrolyte deposition conditions. Nickel oxide, unlike iron oxide, exhibits low solubility in lithium carbonate. However, as with nickel, we expect a wide variation of nickel deposition, with a variation of electrolysis conditions towards the goal of developing nickel as nucleation points to activate carbon nanofiber growth.

As will be demonstrated, even a low concentration of nickel originating from corrosion of the anode can deposit onto the cathode and <u>catalyze carbon nanofiber formation</u>. As illustrated in Fig. 2, higher nickel release during the carbonate electrolysis, which can be controlled due to the proportionality of the amount of nickel released by the size of the anode, results in a proliferation of CNFs of various diameters. As illustrated in Fig. 3, a lower concentration of controlled release of nickel results in the generation of highly homogeneous, smaller diameter CNFs. Both products were formed during electrolysis on a steel wire cathode in 730°C Li$_2$CO$_3$ with added Li$_2$O. The Fig. 2 carbon product was generated in a cell containing a purposefully oversized Ni anode, and the wide variety of carbon nanofibers generated in this single experiment range from ~0.2 to 4 μm in diameter and are up to 100μm in length.

As seen in the electron dispersive spectroscopy in Fig. 3, the bright spots are Ni deposits on the cathode. This is consistent with the known alternative CVD CNF growth mechanism, in which Ni acts as nucleation sites to initiate CNF growth[16]. The XRD diffraction peaks at 26° and 43° are assigned to the hexagonal graphite (002) and diffraction planes (JCPDS card files no. 41-1487) within the CNF (specifically, the stacking of parallel graphene layers and the size of graphene layer, respectively)[10].

The predominant CNF cathode product is observed when the electrolysis is initiated at low current typically, 5 mA cm$^{-2}$, followed by an extended high current electrolysis such as at 100 mA cm$^{-2}$. The cathode product is principally amorphous (and only ~25% CNF), when staring directly at only a high (100 mA cm$^{-2}$) current density. We interpret this mechanistically as follows: due to its low solubility, and lower reduction potential, nickel (in this case originating from the anode) is preferentially deposited at low applied electrolysis currents (5 or 10 mA cm$^{-2}$). This is evidenced by the low observed electrolysis voltage (<0.7V) and sustains the formation of nickel metal cathode deposits, which appear to be necessary to nucleate CNF formation. The concentration of electrolytic [$CO_3^{2-}$] >> [$Ni^{2+}$], and mass diffusion dictates that higher currents will be dominated by carbonate reduction. The subsequent higher electrolysis voltage thermodynamically required to deposit carbon[23] is only observed at higher applied currents (> 20 mA cm$^{-2}$). Hence, without the initial application of low current, amorphous carbon will tend to form, while CNFs are readily formed following the low current nickel nucleation activation. In our typical high yield CNF electrolyses, the current is increased from 50 to 1000 mA at a 10 cm$^{-2}$ steel cathode over 15 minutes, followed by a constant current electrolysis at 1000 mA.

The CNF's synthesized in Fig. 3 utilized the same cathode and same composition electrolyte as in the Fig. 2 synthesis, but with a controlled, limited amount of nickel released to limit the Ni nucleation points forming on the cathode. This was accomplished by using a smaller (by one third) surface area of nickel anode in the electrolysis cell. The better resolved XRD peaks at 43° (100 plane) & 44° (101 plane) in Fig. 3 compared to Fig. 2 is evidence of better homogeneity of the CNFs synthesized under the Ni-limited conditions. As illustrated in Fig. 3, a lower concentration of controlled release of nickel results in the generation of highly homogeneous CNFs with a smaller, consistent diameter of ~0.2 μm.

The high nickel electrolysis product examined in the SEM of Fig. 2 present a proliferation of tangled carbon nanofibers of a wide variety of diameters. The low nickel electrolysis product examined in the SEM of Fig. 2 and 3 present a tangle of homogeneous-diameter carbon nanofibers. Each of those electrolyses utilized a molten $Li_2CO_3$ electrolyte with 6 m added (dissolved) $Li_2O$. Evidently, high concentrations of oxide localized in the nanofiber formation region leads to torsional effects (tangling). As shown in Fig. 4 carbon nanofibers grown by electrolysis in pure molten $Li_2CO_3$ (without added $Li_2O$) are consistently untangled, uniform and long. The nanofibers range from 300 to 1000 nm in width and 20 to 200 μm long.

The linear EDS map on the middle, right side of Fig. 4 shows elemental variation along the 6μm path of the EDS scan from pure Ni at the start of the fiber to pure C along the remainder of the fiber. Coulombic efficiency percent compares the moles of carbon recovered to applied each 4 moles of electrolysis charge in the reduction of $CO_2$) and is over 80% (and approaches 100% with carefully recovery of all product after washing), the product (after washing off the electrolyte) consists of >80% pure carbon nanofibers. Initial results, indicative of whether the extent to which the nanofibers are hollow (tubes) or filled (threads), is seen in the SEM and TEM at the bottom of Fig. 4. Control of the number of carbon shells comprising the nanofiber, and the extent of filled fibers, and initial indications that Pt and Zn tend not to nucleate molten electrolytic CNF formation, but along with Ni, that Co, Cu and Mg might also induce CNF nucleation, will be presented in ongoing studies.

The CNF electrolysis chamber is readily scalable (Extended Data Fig. 2); we have scaled to 100A using 300 cm$^2$ (shown) and 800 cm$^2$ electrode cells, which scale the smaller cells both in lower carbon splitting potentials (1-1.5 V) and high (80-100 %) 4 electron coulombic efficiency of carbon product formation.

The demonstrated CNF synthesis can be driven by any electric source. As an alternative to conventionally generated electrical, we have also driven the CNF synthesis using electric current as generated by an illuminated efficient concentrator photovoltaic operating at maximum power point. We have previously heated $CO_2$ as a reactant for the electrolysis cell by (i) initially passing it over and under the concentrator photovoltaic (CPV), and then (ii) heating it to the electrolysis temperature using sub-band gap (infrared) thermal light split from concentrated sunlight (via a hot mirror) prior to absorption by the CPV.[6] In this study, to demonstrate the efficient solar synthesis of CNFs, we use an indoor solar simulator, and a 39% solar efficient CPV operating at 550 suns concentration (Extended Data Fig. 3), which has a maximum power point voltage of 2.7 V to drive two in-series (1.35V x 2) CNF electrolyzers at 2.3A. CNFs are efficiently formed from $CO_2$ in air, using inexpensive Ni and steel electrodes at high solar efficiency.

The demonstrated carbon dioxide to CNF process can consist of solar driven and solar thermal assisted $CO_2$ electrolysis. Large scale-up of STEP CNF could initiate with smokestack emissions providing a $CO_2$ reactant which is both hot and concentrated. An even greater amelioration of climate change will occur with the direct removal of atmospheric $CO_2$. Extrapolating the present scale of the solar CNF synthesis determines that 700 $km^2$ of CPV in an area < 10% of that of the Sahara Desert will decrease atmospheric $CO_2$ to pre-industrial concentrations in ten years (Extended Data Schematic 1). Industrial environments provide opportunities to further enhance the $CO_2$ extraction rate; for example fossil-fuelled burner exhaust provides a source of relatively concentrated, hot $CO_2$ requiring less energy than the room temperature dilute $CO_2$ in the atmosphere. The product, carbon nanofiber may be stored as a stable, dense resource for future generations or stored in widespread use as a flexible, conductive, high strength material in carbon composites for infrastructure, transportation and consumer devices.

It is of interest whether material resources are sufficient to expand the process to substantially impact (decrease) atmospheric levels of carbon dioxide. The build-up of atmospheric $CO_2$ levels from a 280 to ~400 ppm occurring over the industrial revolution comprises an increase of 2 x $10^{16}$ mole (8.2 x$10^{11}$ metric tons) of $CO_2$, and will take a comparable effort to remove. It would be preferable if this effort results in useable, rather than sequestered, resources. We calculate below a scaled up capture process can remove all excess atmospheric $CO_2$ converting it to useful carbon nanofibers.

Via the Faraday equivalence approaching 100% coulombic efficiency, 0.3 A $cm^{-2}$ will remove 10 tonnes of carbon dioxide per $m^2$ cathode per year, calculated as: 3x$10^3$ A $m^{-2}$ x 3.156x$10^7$ s $year^{-1}$ x (1 mol $e^-$ / 96485 As) x ($CO_2$/ 4 $e^-$)) x (4.40098 x$10^{-6}$ tonne / mol) = 10 tonne $CO_2$. At full absorption and conversion of $CO_2$, this would require air with 0.04% $CO_2$ striking the cell with a wind speed of 1 mph per as: 1609 m air per h x 1 $m^2$ x 8,766 hour per year) x (0.0004 $m^3$ $CO_2$ / $m^3$ air) x 1 tonne per 556 $m^3$ $CO_2$ = 10 tonne $CO_2$.

In STEP (solar thermal electrochemical process) CNF capture, 6 kWh $m^{-2}$ of sunlight per day, at 500 suns on 1 $m^2$ of 39% efficient CPV, will generate 430 kAh at 2.7 V to drive two series connected series connected molten carbonate electrolysis cells to form carbon nanofibers. As summarized in Extended Data Schematic 1, this will capture 8.1 x$10^3$ moles of $CO_2$ $day^{-1}$ to form solid carbon (based on 430 kAh · 2 series cells / 4 Faraday $mol^{-1}$ $CO_2$). The material resources to decrease atmospheric carbon dioxide concentrations with STEP CNF capture, appear to be reasonable. From the daily conversion rate of $8.1 \times 10^3$ moles of $CO_2$ per square meter of CPV, the capture process, scaled to 700 $km^2$ of CPV operating for 10 years can remove and convert all the increase of $2 \times 10^{16}$ mole of atmospheric $CO_2$ to solid carbon. A larger current density at the electrolysis electrodes, will increase the required voltage and would increase the required area of CPVs. A variety of CSP installations, which include molten salt heat storage, are being commercialized, to permit 24/7, rather than daylight only, operation, and costs are decreasing. STEP provides higher solar energy conversion efficiencies than CSP, by producing a chemical product rather than electricity, and secondary losses can be lower (for example, there are no grid-related transmission losses). Contemporary concentrators, such as based on plastic Fresnel, rather than power towers have shorter focal length to avoid environmental thermal damage. Heat exchange losses are to be expected between hot, $CO_2$-cleansed air and ambient input air. A greater degree of solar concentration, for example 2000 suns, rather than 500 suns, will proportionally decrease the quantity of required CPV to 175 $km^2$, while the concentrator land area will be several thousand fold higher than the CPV area, equivalent to < 10% of the area of the Sahara desert (which averages ~6 kWh $m^{-2}$ of daily sunlight over its $10^7$ $km^2$ surface), to remove anthropogenic carbon dioxide in ten years.

A related resource question is whether there is sufficient lithium carbonate, as an electrolyte of choice to decrease atmospheric levels of carbon dioxide for the STEP CNF carbon capture. 700 $km^2$ of CPV plant will require several million tonnes of lithium carbonate electrolyte, depending on the electrolysis cell thickness current density and cell thickness. Thicker, or lower current density, cells will require proportionally more electrolyte. Fifty, rather than ten, years to return the atmosphere to pre-industrial carbon dioxide levels will require proportionally less electrolyte. These values are viable within the current production of lithium carbonate. Lithium carbonate availability as a global resource has been under scrutiny to meet the growing lithium battery market. It has been estimated that the annual production will increase to 0.24 million tonnes by 2015. Sodium and potassium carbonate are substantially more available and our study of these as alternate or mixed electrolytes is in progress.

Here, we show a highly valued, stable, compact carbon, carbon nanofiber, that can be directly formed from atmospheric or exhaust $CO_2$ in an inexpensive process. Today, CNFs require 30 to 100 fold higher production energy compared to aluminum. We present the first high yield, inexpensive synthesis of carbon nanofibers from the direct electrolytic conversion of $CO_2$, dissolved in molten carbonates to CNFs at high rates using scalable, inexpensive nickel and steel electrodes. The structure is tuned by controlling the electrolysis conditions, such as the addition of trace nickel to act as CNF nucleation sites, limits to the electrolytic oxide concentration, and control of current density. New infrastructure and merchandise built from CNFs would provide a massive repository to store atmospheric $CO_2$.

**Methods**

**Reagents.** Barium carbonate (Alfa Aesar, 99.5%), lithium carbonate (Alfa Aesar, 99%), lithium oxide (Alfa Aesar, 99.5%), and calcium carbonate (Alfa Aesar, 98%) are combined to form various molten electrolytes.

**Electrolyses.** Electrolyses are driven at a 2.3 A (amp) at the maximum power point of the illuminated concentrator photovoltaic as shown in Fig. 7, or galvanostatically at a set constant current (constant current) as described in the text. The electrolysis is contained in a pure alumina (AdValue, 99.6%) crucible or pure nickel crucible (Alfa Aesar). Alumina crucible electrolyses used coiled Ni wire (Alfa Aesar, 99.5%) as the (oxygen generating) anode or in scale-up experiments cylinders formed from pure Ni shim (McMaster 9707K5), while electrolyses in the Ni crucibles used the inner walls of the crucible as the anode. A wide variety of steel wires for coiled cathodes are effective, an economic form (used in this study) is Fi-Shock 14 Gauge, Steel Wire model #BWC-14200. During electrolysis, the carbon product accumulates at the cathode, which is subsequently removed and cooled. Details of solar (STEP methodology) electrolyses are provided in references 5, 6 and 28.

**Characterization.** The carbon product is washed, and analyzed by PHENOM Pro-X Energy Dispersive Spectroscopy on the PHENOM Pro-X SEM; by XRD analysis conducted at a sweep rate of 0.12 degree per minute on a Rigaku Miniflex diffractometer with a 0.01 degree slit width, analyzed using the Jade software package (JADE, 6:1; Materials Data, Inc. Livermore, CA, 2002); and by TEM, measured with a JOEL JEM-1200 EX Transmission Electron Microscope.

**Scale-up.** The STEP CNF electrolysis chamber is readily scalable; we have scaled to 100A using 300 (Extended data Fig. 2) and 800 $cm^2$ electrode cells driven with a Xantrex XTR 0-100 A DC power supply, which scale the smaller cells both in lower carbon splitting potentials (1-1.5 V) and high (80-100 %) 4 electron coulombic efficiency of carbon product formation. The larger cells use a cylindrical stainless steel 316 cylinder as a cathode sandwiched within concentric, cylindrical nickel anode anodes.

**Solar Electrolysis.** The process is demonstrated as a scaled-up stand-alone electrolytic cell, and is also shown to be with high solar to electric efficiency to drive the electrolytic CNF production. In the latter case, the electrolyzer current is provided by a 39% solar to electric efficient concentrator photovoltaic (0.3074 $cm^2$ Envoltek ESRD055 CPV) in lab under 1 kW Xenon, daylight color (5600K) AM1(air mass) illumination focused by Fresnel lens to to 550 suns concentration. As shown in Extended Data Fig. 3, the CPV is situated under the AM1 filter, with a fresnel concentrator above the AM 1 light source.

**Figure Legends**

**Figure 1. $CO_2$ to carbon without carbon nanofibers in zero nickel environment.** Top middle, top right: typical maximum variation of observed cathodes subsequent to removal from carbonate electrolytes after a longer (4Ah) electrolysis in molten carbonate. Left photo shows 10cm$^2$ coiled steel wire (0.12 cm diameter) cathode prior to electrolysis. SEM of the washed cathode product subsequent to a <u>nickel-free</u>, 1.5 hour, 1 A constant current electrolysis in 730°C molten $Li_2CO_3$ with 6 m $Li_2O$ using a 10cm$^2$ Pt foil anode and a 10cm$^2$ coiled steel wire (0.12 cm diameter) cathode, in a Ni-free crucible.

**Figure 2. $CO_2$ to a diverse range of carbon nanofibers from high nickel media** (subsequent to electrolysis with an oversized nickel anode, the inner wall of a 50 ml Ni crucible, at 0.05 A, then 1 A constant current electrolysis for 2 hours in 730°C molten $Li_2CO_3$ with 6 m $Li_2O$ using a 10cm$^2$ coiled, 0.12 cm diameter steel wire). Top: X-ray powder diffraction of the carbon product. Middle, bottom: SEM of the washed cathode product subsequent to electrolysis with an oversized nickel cathode (the inner wall of a 50 ml Ni crucible. Consistent tangled fiber product with a wide range of diameter are observed throughout. Typical constant current electrolysis potentials range from 0.5 to1.5 V as the current density is increased from 0.05 A to 1A.

**Figure 3. $CO_2$ to homogenous set of carbon nanofibers from controlled nickel media.** $CO_2$ to homogenous set of carbon nanofibers from controlled nickel media, 6 m added $Li_2O$ electrolyte. Left, middle: (repeated, representative of) EDS on fiber showing high carbon concentration; EDS Left, bottom: (repeated, representative) EDS on bright spot showing high nickel concentration. Right, bottom: SEM of washed cathode product subsequent to electrolysis with a smaller nickel cathode (the inner wall of a 20 ml Ni crucible), 0.05 A, then 1 A constant current electrolysis in 730°C molten $Li_2CO_3$ with 6 m $Li_2O$ using a 10cm$^2$ coiled steel wire (0.12 cm diameter). Throughout the sample SEM consistently exhibited the same diameter CNFs. Top: X-ray powder diffraction of the cathode carbon product.

**Figure 4. $CO_2$ to homogenous, untangled set of carbon nanofibers from controlled nickel media, 0 added $Li_2O$ electrolyte.** SEM of the washed cathode product subsequent to electrolysis with a smaller nickel cathode (the inner wall of a 20 ml Ni crucible), 0.05 A, then 1 A constant current electrolysis in 730°C molten $Li_2CO_3$ using a 10cm$^2$ coiled steel wire (0.12 cm diameter). Middle right: EDS composition mapping along the 6μ path, blue arrow path shown in the middle, left SEM. Lower right: TEM of fiber tip.

**Extended Data Figure 1. Fe variation with different electrolysis conditions.** Iron formed from 1.5 Ah of charge between a 10cm$^2$ Ni anode and a 10cm$^2$ iron cathode, at respective currents (from left to right and top to bottom) of 0.1A, 0.5 A, 2.5 A, 5A, 5A and 5A. In the first 5 panels the electrolysis was conducted in 730°C $Li_2CO_3$ with 3 m $Li_2O$ and 1.5m $Fe_2O_3$, and in the last (lower right) panel 650°C $Li_{1.6}Ba_{0.3}Ca_{0.1}CO_3$ with 6 m LiOH and 1.5m $Fe_2O_3$. Subsequent to water/dilute acid washing to remove excess electrolyte, EDS and titration[28] analysis distinguishes the amorphous particles as iron metal and the octahedra as $Fe_3O_4$. The presence of water available from the equiibrium

$2LiOH \rightleftharpoons Li_2O + H_2O$ in the latter electrolyte encourages the oxidation of deposited iron and formation of a product of octrahedral magnetite, $Fe_3O_4$.

**Extended Data Figure 2. Scaled-up electrolysis chamber splitting $CO_2$ operating at 100 amp, 1.5V, and 95 to 100% coulombic efficiency of the 4 electron reduction to carbon.** Anode and cathode area 300 $cm^2$, alumina cell diameter 7 cm, height 14 cm. Top, left top: Concentric nickel anodes (electrically connected at bottom) with high current bus bar. Top, middle: steel cathode with high current bus prior to insertion between concentric anodes. Top, right: full cell prior to addition of the carbonate electrolyte. Bottom, left: Cathode subsequent to 1 hour, 100 A electrolysis at 1.5V with thick carbon deposited coating. Bottom, middle: Steel removal of carbon by striking cathode with hammer, and (view from bottom up) cathode with carbon layer removal nearly complete. Bottom, right: 2nd electrolysis with same electrodes yields same results.

**Extended Data Figure 3. Concentrator photovoltaic driving STEP CNF synthesis.** Electrolyzer current by power supply (initial experiments) and now provided by an efficient concentrator photovoltaic in lab under 1 kW Xenon, daylight color (5600K) AM1(air mass) illumination. Left side is the 0.3074 $cm^2$ Envoltek ESRD055 CPV situated under the air-cooled AM1 filter. Middle top: The Fresnel concentrator above the AM 1 filter. Middle bottom: the unattached CPC under the secondary optical concentrator. Right side: Typical (550 sun) photocurrent - voltage plot of the CPV.

**Extended Data Schematic 1. STEP carbon nanofiber to lower atmospheric $CO_2$ to preindustrial levels in 10 years.** The required CPV and land area updated from 2010 STEP carbon and 2010 $CO_2$ concentration calculations in reference 23.

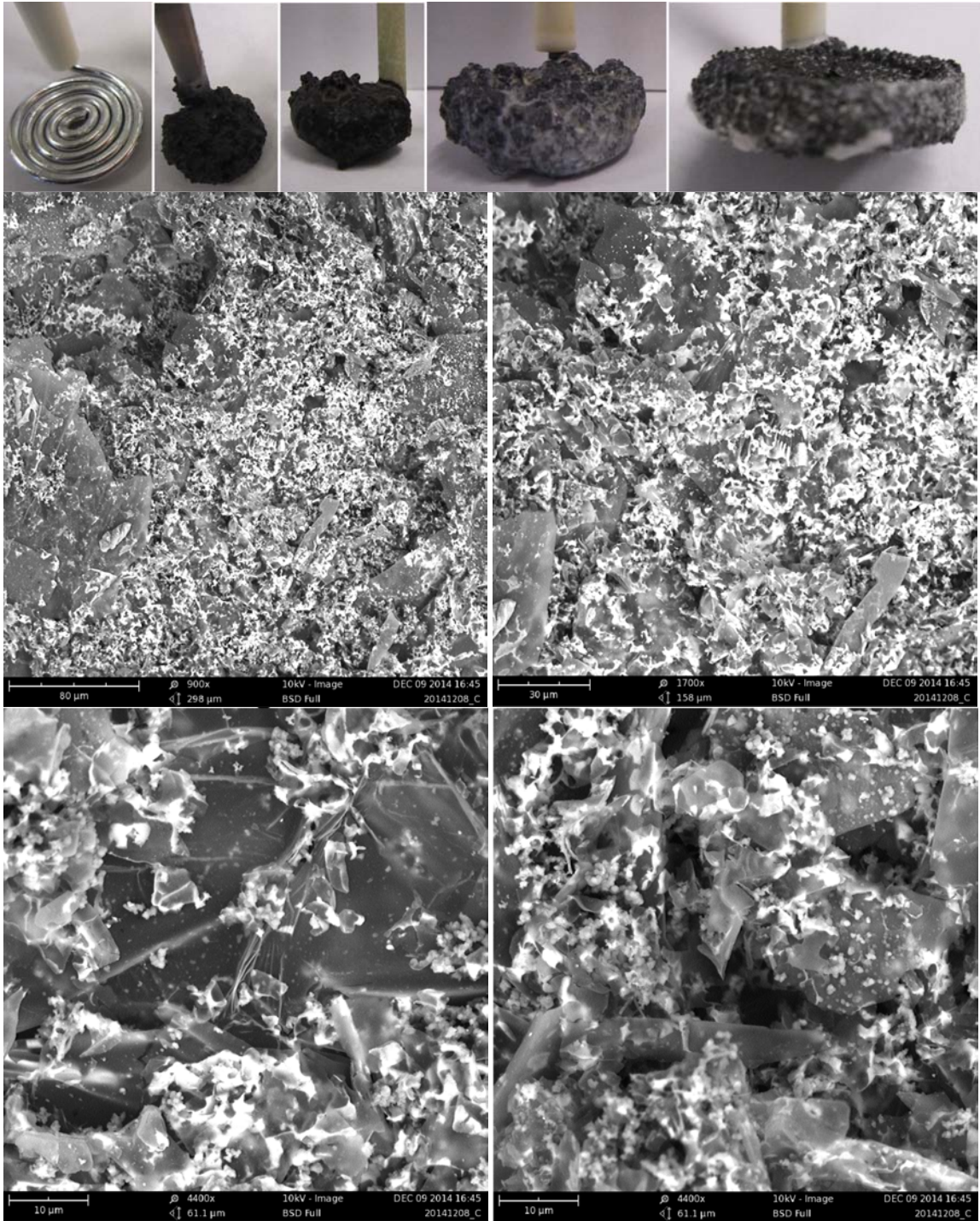
**Figure 1. CO₂ to carbon without carbon nanofibers in no nickel environment.**

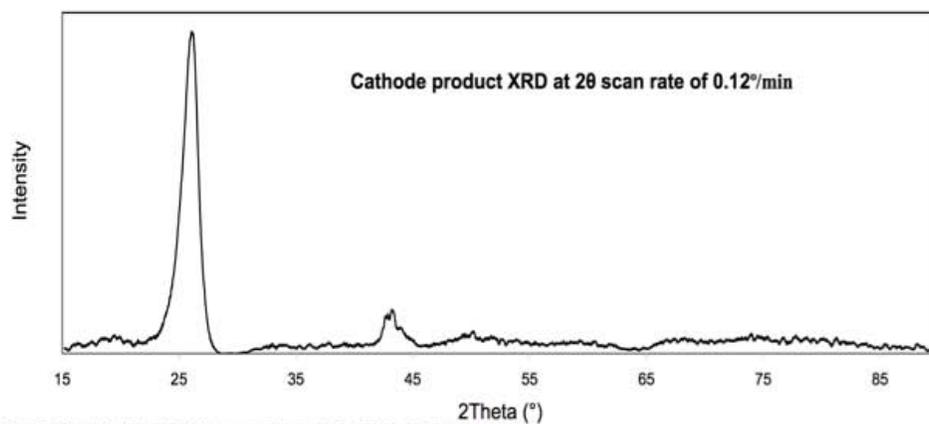
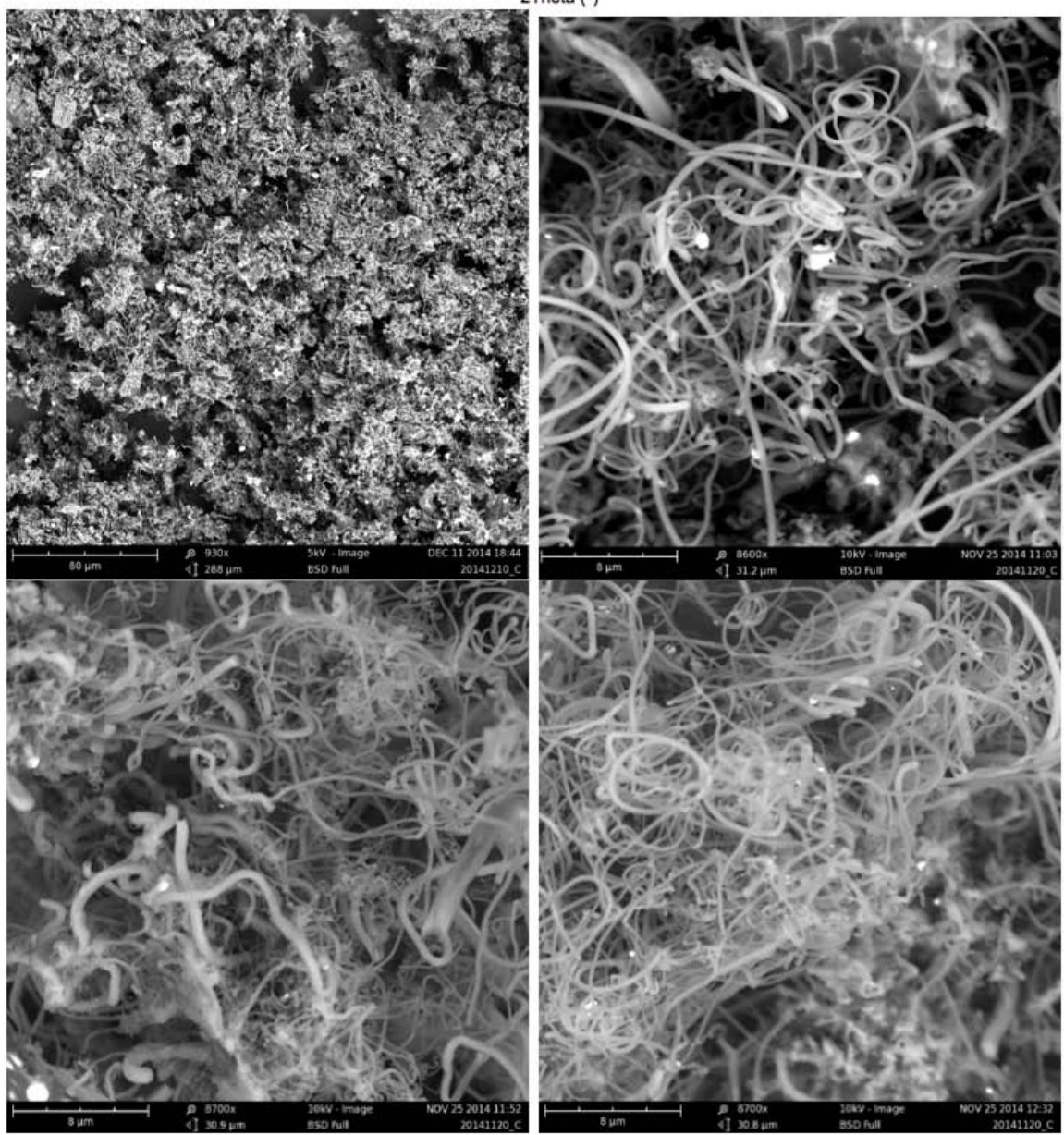

**Figure 2. CO₂ to a diverse range of carbon nanofibers from high nickel media.**

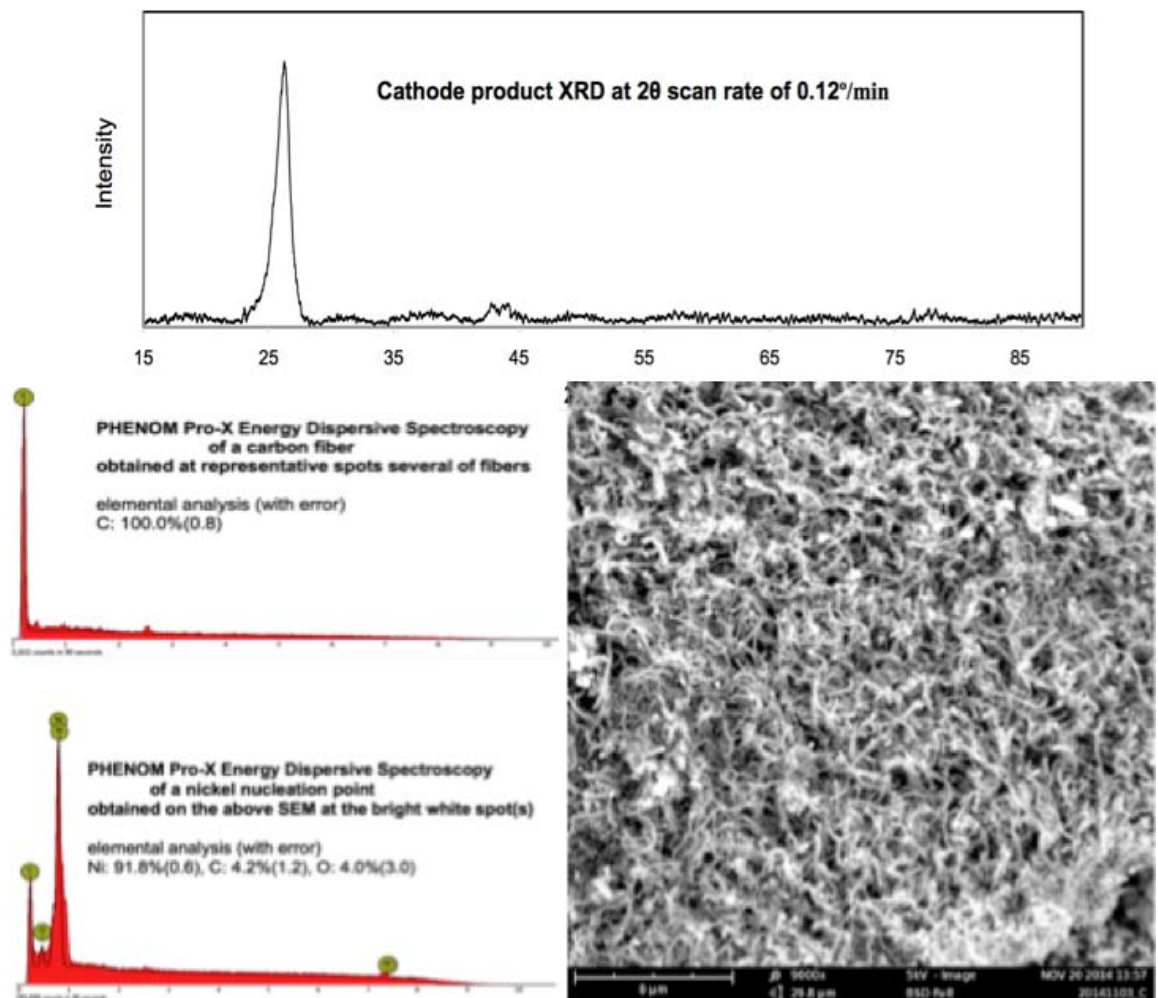

**Figure 3. CO$_2$ to homogenous set of carbon nanofibers from controlled nickel media.**

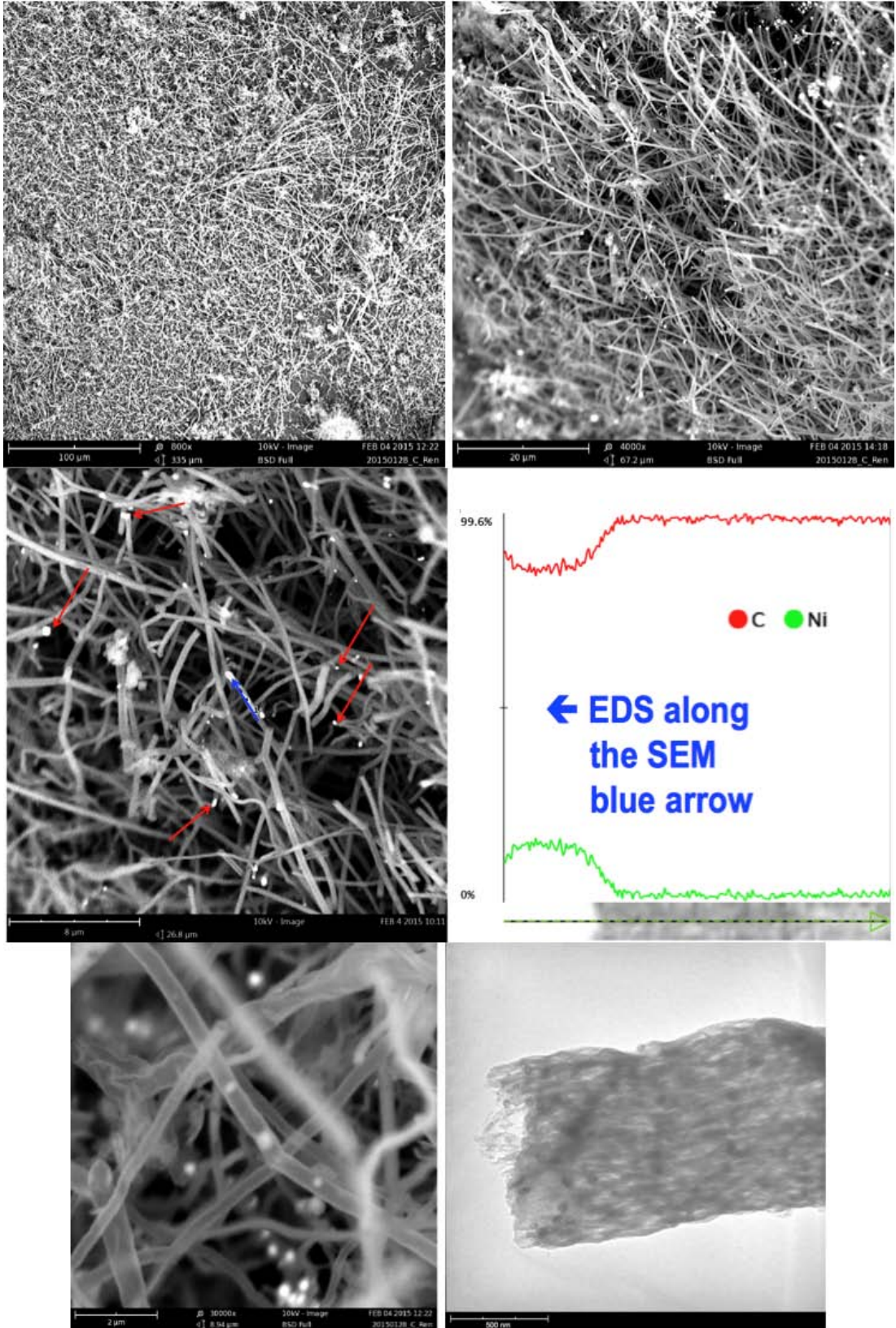

**Figure 4. CO$_2$ to homogenous, untangled set of carbon nanofibers from controlled nickel media, 0 added Li$_2$O electrolyte.**

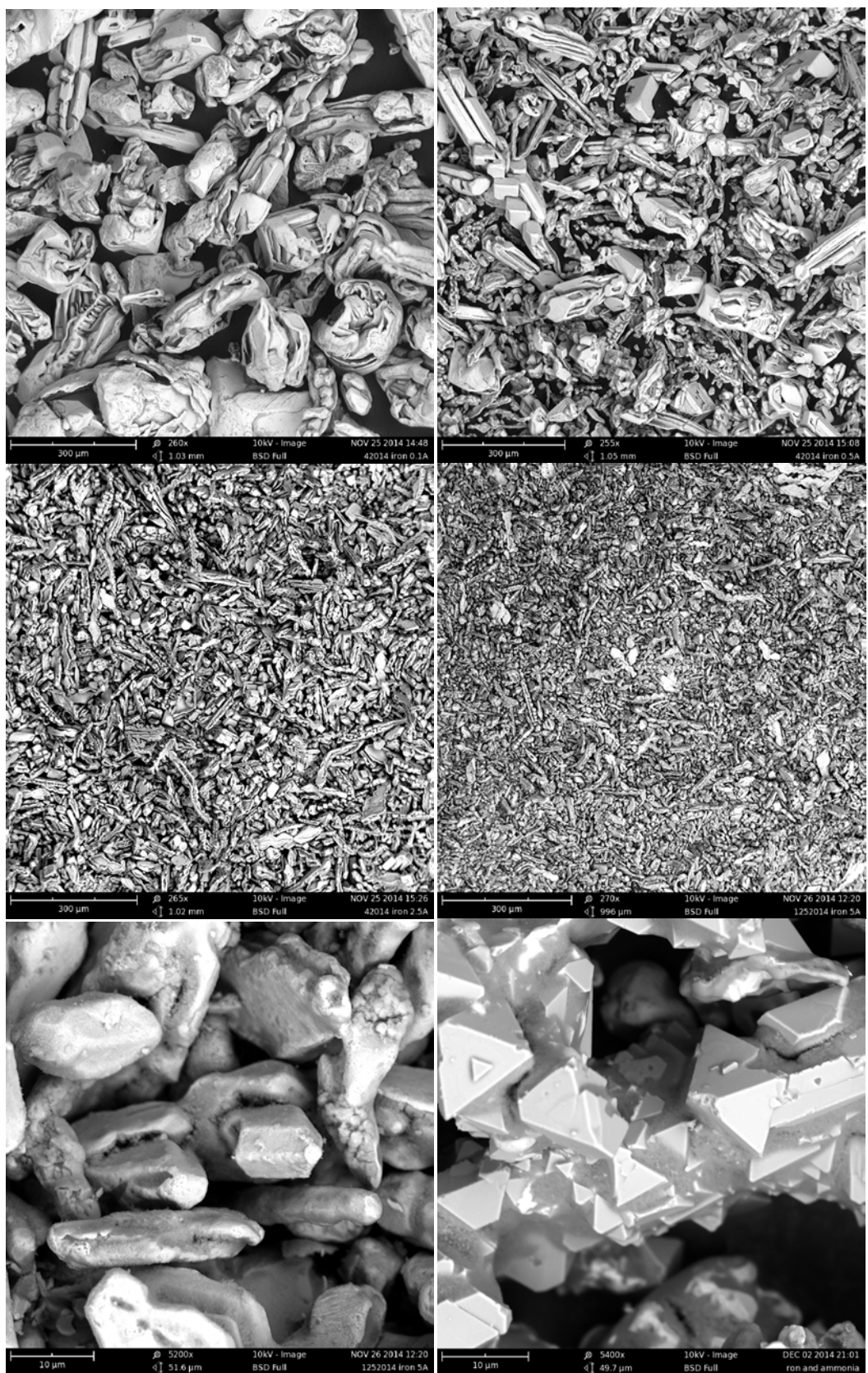

**Extended Data Figure 1. Fe variation with different electrolysis conditions.**

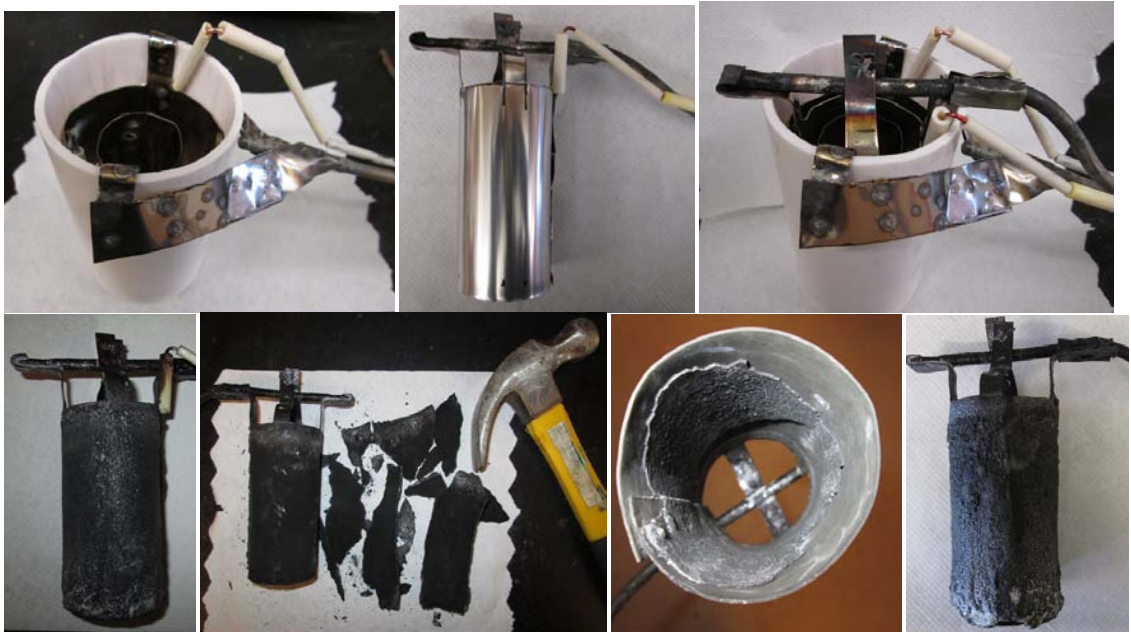

**Extended Data Figure 2. Scaled-up electrolysis chamber splitting $CO_2$ operating at 100 amp, 1.5V, and 95 to 100% coulombic efficiency of the 4 electron reduction to carbon.**

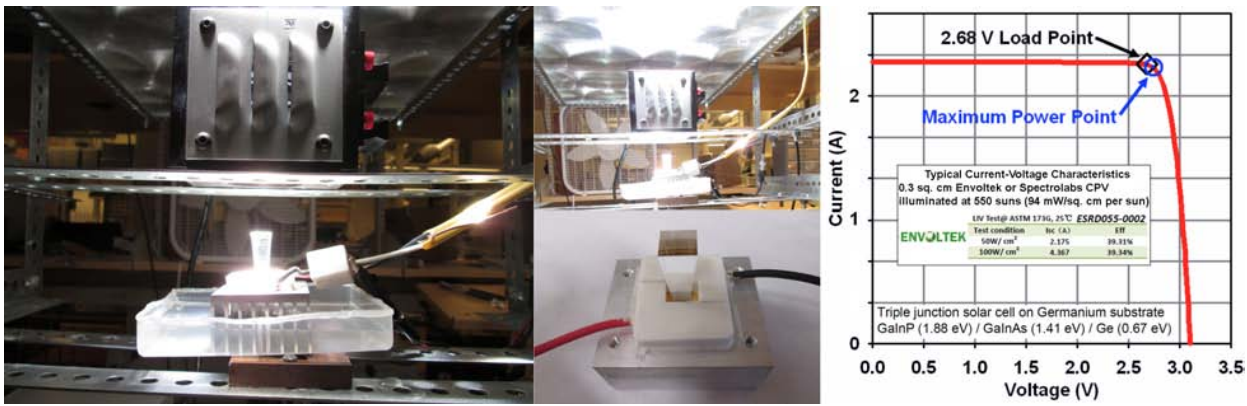

**Extended Data Figure 3. Concentrator photovoltaic driving STEP CNF synthesis.**

> **Are resources sufficient to expand STEP CNF process to decrease atmospheric $CO_2$?**
>
> ♦ During the industrial revolution $CO_2$ has risen from 280 to 397 ppm, a **$2 \times 10^{16}$ mole $CO_2$ increase**.
>
> 6 kWh m$^{-2}$ of sunlight per day, at 500 suns focused on 1 m$^2$ of 39% efficient CPV, will generate 430 kAh at 2.7 V max power to drive 2 series (1.35V each) connected molten carbonate electrolysis cells.
>
> The demonstrated STEP solar efficiency of $CO_2$ conversion is > 39%, as it also incorporates waste CPV solar heat to heat the electrolysis. The challenge is to effectively direct this solar waste heat to $CO_2$ heating[23]: for example recovery from air heated rather than smoke-stack heated $CO_2$ converted to CNFs.
>
> Daily, 430 kAh per day · 2 series electrolysis cells / 4 Faraday mol$^{-1}$ $CO_2$, will capture (per m$^2$ CPV): $8.1 \times 10^3$ moles of $CO_2$ to form solid carbon nanofibers.
>
> ♦ From the daily conversion rate of $8.1 \times 10^3$ moles of $CO_2$ per day per square meter of CPV, scaled to 700 km$^2$ of CPV **STEP CNF operating for 10 years**: can convert/**decrease all the industry added atmospheric $2 \times 10^{16}$ mole of $CO_2$** back to CNF.
>
> Higher current density, will increase the electrolysis voltage and increase the required area of CPV, while higher solar concentration, will proportionally decrease the required CPV area.
>
> ♦ 700 km$^2$ of CPV will require a sunlight concentration area *< < 10% of the 10$^7$ km$^2$ area of Sahara desert, for STEP CNF removal of anthropogenic carbon dioxide in 10 years*.

**Extended Data Schematic 1. STEP carbon nanofiber to lower atmospheric $CO_2$ to preindustrial levels in 10 years**.